\documentclass[%
  preprint,
superscriptaddress,
 amsmath,amssymb,
 aps,
 floatfix
]{revtex4-1}

\usepackage[dvipdfmx]{graphicx}% Include Figure files
\usepackage{dcolumn}% Align table columns on decimal point
\usepackage{bm}% bold math
\usepackage{color}

\usepackage{lineno}
\newcommand{\di}{\,\mathrm{d}}
\newcommand{\pdiff}[2]{\frac{\partial #1}{\partial #2}}
\newcommand{\im}{\mathrm{i}}

\begin{document}

\title{Low-dimensional dynamics of phase oscillators driven by Cauchy noise}
\author{Takuma \surname{Tanaka}}
\affiliation {Graduate School of Data Science, Shiga University, 1-1-1 Banba, Hikone, Shiga 522-8522, Japan}
\email{tanaka.takuma@gmail.com}
\date{\today}

\begin{abstract}
Phase oscillator systems with global sine-coupling are known to exhibit low-dimensional dynamics.
In this paper, such characteristics are extended to phase oscillator systems driven by Cauchy noise.
The low-dimensional dynamics solution agreed well with the numerical simulations of noise-driven phase oscillators in the present study.
The low-dimensional dynamics of identical oscillators with Cauchy noise coincided with those of heterogeneous oscillators with Cauchy-distributed natural frequencies.
This allows for the study of noise-driven identical oscillator systems through heterogeneous oscillators without noise and vice versa.

\begin{description}
%\item[Usage]
%Secondary publications and information retrieval purposes.
%05.45.Xt Synchronization; coupled oscillators
%05.40.Ca Noise

\item[PACS numbers]
  05.45.Xt, 05.40.Ca
%May be entered using the \verb+\pacs{#1}+ command.
\end{description}
\end{abstract}

\pacs{Valid PACS appear here}

\maketitle

\section{Introduction}
The synchronized rhythmic flashing of fireflies is a spectacular example of a collective phenomenon \cite{Buck1988}.
Fireflies exhibit different and fluctuating flashing frequencies and can be regarded as heterogeneous and noisy oscillators.
Both heterogeneity and noise are essential properties of systems that display collective phenomena.
Coupled phase oscillators have been used to examine how heterogeneity and noise affect the synchronization of physical, chemical, and biological systems \cite{Kuramoto1984,Acebron2005}.
Phase oscillator systems with heterogeneous natural frequencies have been studied since the invention of the phase oscillator model.
Ott and Antonsen \cite{Ott2008} showed that the behavior of globally sine-coupled oscillators, the natural frequencies of which obey a family of rational distribution functions, can be described by low-dimensional dynamics.
%もし自然振動数がコーシーあるいはローレンツ分布に従うならば、なんと無限個の振動子全体のダイナミクスが単一のStuart--Landau方程式で記述されてしまう。自然振動数分布が違った形状だったり結合が単一でない場合には単一ではなくなるが、しばしばcoupled Stuart--Landauで記述でき、ダイナミクスについての多くの情報が得られる。
Specifically, if the natural frequencies obey the Cauchy or Lorentzian distribution, the dynamics of an infinite number of oscillators are described by a Stuart--Landau equation, i.e., a two-dimensional dynamical system.
If the coupling strength takes on several values or the natural frequencies obey the mixture of Cauchy distributions, the dynamics are described by coupled Stuart--Landau oscillators.
%下のコーシーあるいはローレンツは消す
%驚くべきことは、これが何らかの級数展開の低次項をとってきたのではなく、特殊な初期条件を仮定すれば厳密なことである。
This is an exact result for a specific initial condition and not an approximation obtained by ignoring higher order terms.
This type of low-dimensional description has accelerated the study of the heterogeneous oscillator systems \cite{Martens2009,Hong2011}.

However, investigating noise-driven oscillator systems appears to be more challenging than studying heterogeneous oscillator systems.
Previous studies have approximated the dynamics with circular cumulants to obtain low-dimensional dynamics similar to those proposed by Ott and Antonsen \cite{Goldobin2018,Tyulkina2018}.
Although this approach has been implemented with some success, it is not always free from approximation error.
Determining low-dimensional descriptions with fewer approximation errors will be useful in understanding the collective phenomena in various fields, although it may not be as general as approximation with circular cumulants.
This may be possible using a noise that adheres to the assumption of the analysis by Ott and Antonsen.
%これはOtt-Antonsenの記述を出ないようなノイズを選ぶことができれば達成できる可能性がある。

This paper reports that systems driven by Cauchy noise can be described by closed-form low-dimensional dynamical equations.
Phase oscillator systems driven by Cauchy or more generally by non-Gaussian noise have not been studied in as much detail as those driven by Gaussian noise.
However, non-Gaussian noise is known to be prevalent in biological systems \cite{Segev2002}.
For example, a circular auto-regressive model with wrapped Cauchy noise has been proposed to model animals' direction of travel \cite{Shimatani2012}.
Thus, the behavior of phase oscillators driven by Cauchy noise is worthy of further examination.
In addition, as harmonic oscillators display nontrivial phase distribution under L\'{e}vy noise \cite{Sokolov2011}, the dynamics of phase oscillators driven by Cauchy noise is of interest.

This paper is organized as follows.
First, the Watanabe--Strogatz theory is reviewed and used to derive the low-dimensional dynamics of the order parameter of identical sine-coupled oscillators driven by Cauchy noise.
Second, the Ott--Antonsen ansatz is reviewed, and the dynamics of the order parameter of heterogeneous noise-driven oscillators are derived.
It is shown that the amplitude of Cauchy noise and the scale parameter of natural frequency are equivalent in the low-dimensional description, and the implications of the model are discussed.

\section{Analysis and Results}
%まず最初に同一振動子でノイズがない場合を考えます。その後heterogeneityを導入します
This section first considers the system of identical oscillators and then that of heterogeneous oscillators.
Using the notation of Pikovsky and Rosenblum \cite{Pikovsky2015}, we consider the system of $N$ noise-driven phase oscillators with identical natural frequency $\omega$, in which the dynamics of oscillator $k$ are given by

\begin{align}
  \dot{\phi}_k &= \omega+\Im[H(t)\exp(-\im\phi_k)]+\sigma(t)\xi_k(t) \nonumber\\
  &= \omega + |H(t)|\sin[\arg H(t)-\phi_k]+\sigma(t)\xi_k(t), \label{eq:stochastic}
\end{align}
where $H(t)$ is the common forcing, $\sigma(t)>0$ is the amplitude of the noise, and $\xi_k(t)$ is the noise.
%通常はtulkinaのように\xi_k(t)を正規分布に従うものとして考えるが、ここでは違うノイズを考える
This paper uses the Cauchy distribution instead of the Gaussian distribution, which has been used in earlier studies \cite{Goldobin2018,Tyulkina2018}.
It is assumed that $\xi_k(t)$ obeys the independent standard Cauchy distribution without temporal correlation;
%確率密度関数でいう$\xi_k(t)$はこうなる
the probability density function of $\xi_k(t)$ is
%\begin{equation}
%  p[\xi_k(t)]=\frac{1}{\pi\nu}\frac{\nu^2}{[\xi_k(t)-\xi_0]^2+\nu^2}, \label{eq:StandardCauchy}
%\end{equation}
%where the scale parameter $\nu$ is 1 and the location parameter $\xi_0$ is 0.
\begin{equation}
  p[\xi_k(t)]=\frac{1}{\pi}\frac{1}{\xi_k(t)^2+1}. \label{eq:StandardCauchy}
\end{equation}
The common forcing, $H(t)$, can be an external forcing or mutual interaction between oscillators.
For example, the dynamics with $\sigma(t)=\sigma$ and $H(t) = Kz(t)$, where
\begin{equation}
  z(t) = \frac{1}{N}\sum_{k=1}^N \exp(\im\phi_k) \label{eq:orderparameter}
\end{equation}
is the complex-valued order parameter and $K$ is the coupling strength, lead to the following dynamics
\begin{equation}
\dot{\phi}_k = \omega+\frac{K}{N}\sum_{j=1}^N\sin(\phi_j-\phi_k)+\sigma\xi_k(t). \label{eq:Kuramoto}
\end{equation}
In this system, the oscillators are driven by the Cauchy noise and attracted to each other.
%すなわち振動子の重心
%これはノイズ駆動の全結合位相振動子ですね
%数値計算的にはEq.~\ref{eq:stochastic}はタイムステップを\Delta tとしてthe Euler methodでこんな風に
The system of Eq.~(\ref{eq:stochastic}) can be numerically implemented by the Euler method as
\begin{equation}
\phi_k(t+\Delta t) = \phi_k(t)+\Delta t\left\{\omega + |H(t)|\sin[\arg H(t)-\phi_k(t)]+\sigma(t)\xi_k(t)\right\}, \label{eq:Euler}
\end{equation}
where $\xi_k(t)$ follows the standard Cauchy distribution [Eq.~(\ref{eq:StandardCauchy})].
%\sqrt{\Delta t}ではないのはコーシー分布だから
Let us note that the noise term is multiplied by $\Delta t$ instead of $\sqrt{\Delta t}$ because the Cauchy distribution is the stable distribution of index 1.

%これまでの研究で明らかにされたことを論文を引用しながらレビューする
Here, what has been clarified by the previous studies on the behavior of the system without noise is reviewed.
Inserting $\sigma(t)=0$ into Eq.~(\ref{eq:stochastic}) yields
\begin{equation}
  \dot{\phi}_k = \omega+\Im[H(t)\exp(-\im\phi_k)]. \label{eq:deterministic}
\end{equation}
Watanabe and Strogatz \cite{Watanabe1994,Goebel1995} demonstrated that this system is described using three variables and $N-3$ constants of motion.
%the analysis of Watanabe and Strogatzが示すのは、外力を様々に与え、初期にある位相にあった振動子がどこに来るかが、三つのパラメータの写像によって表現できることである
More specifically, the phases $\phi_k(t)\;(1\le k\le N)$ of oscillators driven by the common forcing $H(t)$ are given by a three-parameter function of the initial phases, $\phi_k(0)$.
%Watanabe--Strogatzの考え方に従うと，初期に一様分布に従う無限個の振動子があった場合、その時間発展を記述する三つのパラメータのうちの二つをオーダーパラメタの実数成分と複素数成分とすることができる
%このとき、もう一つのパラメータは初期の一様分布の回転と見なすことができる
Using the Watanabe--Strogatz theory, the function that maps $\phi_k(0)$ to $\phi_k(t)$ is defined by the real and imaginary components of the order parameter and a parameter corresponding to the rotation of the initial phases \cite{Watanabe1994,Pikovsky2008}.
%すなわち、現在の振動子の位相の分布はオーダーパラメータのみによって記述できることになる
%これが何らかの高次の寄与を無視した結果によるものではなく、閉じた記述であることに注意する必要がある
This allows us to obtain a closed-form description of the dynamics of order parameter.
In the following analysis, it is assumed that the constants of motion are uniformly distributed in the limit of an infinite number of oscillators.
%実際には必ずしも一様分布するわけではないが、ランダムな初期状態から出発したときにこの仮定に基づいて計算した結果と一致している。また、非一様な自然振動数分布の場合のN∞の結果はこれできっちり再現できる
This assumption has successfully described the behavior of the finite number of phase oscillators whose initial phases are drawn from the uniform distribution on $[0,\;2\pi]$.
The order parameter $z(t)$ becomes
\begin{equation}
  Z(\omega,t) = \int_0^{2\pi}p(\phi,t|\omega)\exp(\im\phi)\di\phi
\end{equation}
in the limit of $N\rightarrow\infty$, where $p(\phi,t|\omega)$ is the density of the phases of oscillators with natural frequency $\omega$ at time $t$.
%さらにこの場合、オーダーパラメタの時間発展は以下のように書き表せる。
For the system of Eq.~\ref{eq:deterministic}, the dynamics of the order parameter have been shown to follow
\begin{equation}
  \pdiff{Z(\omega,t)}{t} = \im\omega Z(\omega,t)+\frac{H(t)}{2}-\frac{\bar{H}(t)}{2}Z(\omega,t)^2 \label{eq:continuousWS}
\end{equation}
\cite{Pikovsky2008,Marvel2009}.
%委譲のことから、Watanabe-Strogatzで初期に一様分布でありオーダーパラメタによって写像が与えられるならば、振動子の位相分布はオーダーパラメタのみによって与えられる分布になる
Because, if the initial phases are uniformly distributed, the rotation of the initial phase does not affect the final distribution of the phases, the phase distribution of oscillators at $t$ is determined solely by the order parameter \cite{Pikovsky2008}.
%それがこのポアソンカーネルである（引用）
Thus, it has been shown that the density of the oscillators' phase obeys the Poisson kernel \cite{Ott2008}
\begin{equation}
  p(\phi,t|\omega) = \frac{1}{2\pi}\frac{1-|Z(\omega,t)|^2}{1-2|Z(\omega,t)|\cos[\phi-\arg Z(\omega,t)]+|Z(\omega,t)|^2}. \label{eq:poisson}
\end{equation}

%以上でレビューを終えて、本論文で導入するノイズについて考察する。
Having reviewed the previous results, we are prepared to examine the dynamics of noise-driven oscillators.
Because the phase distribution in the system without noise, which has a low-dimensional description, is determined by the order parameter, the system with noise can have a low-dimensional description if the phase distribution is determined by a few parameters.
To obtain a low-dimensional description, it is useful to note that the Poisson kernel is identical to the wrapped Cauchy distribution \cite{Mardia2009}

\begin{align}
  p(\phi) &= \sum_{n=-\infty}^\infty\frac{\lambda}{\pi[\lambda^2+(\phi-\mu+2\pi n)^2]} \nonumber \\
  &= \frac{1}{2\pi}\frac{\sinh\lambda}{\cosh\lambda-\cos(\phi-\mu)} \label{eq:wrapped}
\end{align}
if the following is set:

\begin{align}
  \mu&=\arg Z(\omega,t), \label{eq:mu}\\
  \lambda&=\sinh^{-1}\left(\frac{|Z(\omega,t)|^{-1}-|Z(\omega,t)|}{2}\right)\nonumber\\
  &=-\log|Z(\omega,t)|. \label{eq:lambda}
\end{align}
Before considering noise-driven sine-coupled oscillators, uncoupled oscillator systems driven by Cauchy noise are examined, that is, $\sigma(t)>0$ and $H(t)=0$.
In this system, assuming that the oscillators are initially distributed according to the wrapped Cauchy distribution [Eq.~(\ref{eq:wrapped})], the Cauchy noise ensures that the oscillators obey the wrapped Cauchy distribution.
%This is because the summation of the random variables that obey the Cauchy distributions with the scale parameters $\lambda_i\;(1\le i\le n)$ obeys the Cauchy distribution with the scale parameter $\sum_{i=1}^n \lambda_i$ for any integer $n$.
%これはなぜなのかを示すためにThe Euler methodで計算することを考える。このとき、
%である。
This is clarified by the Euler method
\begin{equation}
\phi_k(t+\Delta t)=\phi_k(t)+\Delta t\sigma(t)\xi_k(t).
\end{equation}
%このとき，$\phi_k(t_0)$がscale parameter $lambda$のコーシー分布に従っていたとすると，$\phi_k(t_0+n\Delta t)$, where $n>0$, はscale parameter $\lambda+\sigma \Delta t n$のコーシー分布に従う．コーシー分布再生性より。
If $\phi_k(t_0)$ obeys the Cauchy distribution with the scale parameter $\lambda$ and the location parameter $\mu$, $\phi_k(t_0+n\Delta t)$, where $n>0$, obeys the Cauchy distribution with the scale parameter $\lambda+\Delta t\sum_{j=0}^{n-1}\sigma(t_0+j\Delta t) $ and the location parameter $\mu$ owing to the reproductive property.
Therefore, the Cauchy noise $\xi_k(t)$ increases the scale parameter $\lambda$ as
\begin{equation}
  \dot\lambda = \sigma(t),
\end{equation}
while keeping the location parameter constaint as
\begin{equation}
  \dot\mu = 0.
\end{equation}
Inserting Eqs.~(\ref{eq:mu}) and (\ref{eq:lambda}) gives
\begin{equation}
  \pdiff{Z(\omega,t)}{t} = -\sigma(t) Z(\omega,t). \label{eq:dZdt}
\end{equation}
This equation means that the Cauchy noise causes the exponential decay of the order parameter.

% Eq.~\ref{eq:continuousWS}の記述もEq.~\ref{eq:dZdt}の記述もN無限大なら近似を含まずZについて閉じた記述である。従って、この二つを組み合わせてもZについて閉じた記述になる。そこでこの二つを組み合わせられる。
Because Eqs.~(\ref{eq:continuousWS}) and (\ref{eq:dZdt}) are exact closed-form descriptions of $Z(\omega,t)$ in the limit of $N\rightarrow\infty$, we can combine these two equations to obtain the dynamics of Cauchy noise-driven coupled oscillators.
This is justified by the fact that, the oscillators obeying a wrapped Cauchy distribution remain obeying a wrapped Cauchy distribution if driven by either the sine-coupling or Cauchy noise.
Combining Eq.~(\ref{eq:continuousWS}) with Eq.~(\ref{eq:dZdt}) yields the dynamics of the system with $H(t)\neq 0$ and $\sigma(t)>0$,
\begin{equation}
  \pdiff{Z(\omega,t)}{t} = [\im\omega-\sigma(t)] Z(\omega,t)+\frac{H(t)}{2}-\frac{\bar{H}(t)}{2}Z(\omega,t)^2. \label{eq:Z}
\end{equation}
% きわめて短い時間ごとにノイズと相互作用を切り替えても元の方程式の定義するダイナミクスと同じになる。これはきわめて短い時間で二つの力学系を切り替えた場合に対応するのでこのようになる。
This is equivalent to the system of oscillators driven alternately by Eq.~(\ref{eq:continuousWS}) and Eq.~(\ref{eq:dZdt}).
For globally sine-coupled phase oscillator systems [Eq.~(\ref{eq:Kuramoto})], the common forcing of the oscillators is proportional to the order parameter, that is, $H(t)=KZ(\omega,t)$.
Hence, Eq.~(\ref{eq:Z}) can be used to describe the dynamics of the order parameter of the system of Eq.~(\ref{eq:Kuramoto}) with
\begin{equation}
  \pdiff{Z(\omega,t)}{t} = \left(\im\omega-\sigma+\frac{K}{2}\right)Z(\omega,t)-\frac{K}{2}\bar{Z}(\omega,t)Z(\omega,t)^2. \label{eq:SL1}
\end{equation}
It has a closed-form stable solution
\begin{equation}
  Z(\omega,t) = \begin{cases}
    \sqrt{1-\frac{2\sigma}{K}}\exp[\im\omega (t-t_0)] & (\sigma\le\frac{K}{2})\\
    0 & \left(\frac{K}{2}<\sigma\right)
  \end{cases}, \label{eq:theoretical}
\end{equation}
where $t_0$ is a constant.
%これが意味しているのは、ノイズが弱ければ同期しているが、ノイズの強度sigmaがある閾値を上回ると同期できなくなること
This means that a weak noise allows for the synchronization whereas noise stronger than a threshold value abolishes the synchronization.

\begin{figure}
  \centering\includegraphics[width=8cm]{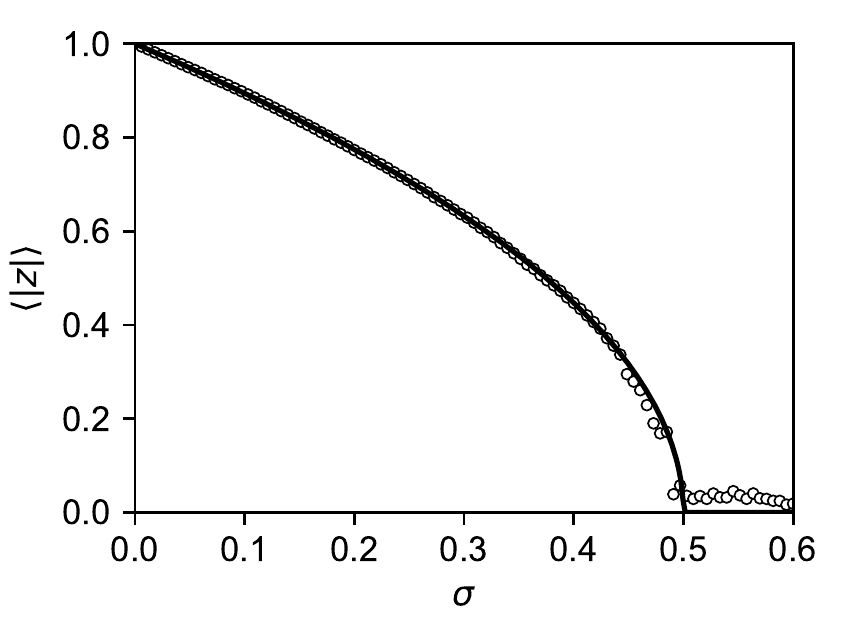}
  \caption{\label{fig:synchronization}
    Numerical and theoretical results of $\langle|z|\rangle$ for the system of Eq.~(\ref{eq:Kuramoto}) with $K=1$.
    The numerical and theoretical results are indicated by the circles and the solid line, respectively.
  }
\end{figure}

To numerically confirm the above theoretical prediction, the simulation of Eq.~(\ref{eq:Kuramoto}) was performed by the Euler method
\begin{equation}
\phi_k(t+\Delta t) = \phi_k(t)+\Delta t\left(\omega + \frac{K}{N}\sum_{j=1}^N \sin[\phi_j(t)-\phi_k(t)]+\sigma\xi_k(t)\right)
\end{equation}
with the parameter values $N=10\,000$, $\omega=0$, and $K=1$ and the simulation time step $\Delta t=0.005$.
%オイラー法で
%であり，\xi_k(t)はindependent standart Cauchy distribution [\Eq.~\ref{eq:StandardCauchy}]からとった。
The noise $\xi_k(t)$ was drawn from the independent standard Cauchy distribution [Eq.~(\ref{eq:StandardCauchy})].
The initial phase was uniformly distributed on $[0,\;2\pi]$.
The average $\langle|z|\rangle$ of the absolute value of the order parameter [Eq.~(\ref{eq:orderparameter})] was obtained during $100\le t\le 200$.
Figure~\ref{fig:synchronization} shows the numerical results of $\langle|z|\rangle$ (circles) and the theoretical value of $|Z(\omega,t)|$ [solid line, Eq.~(\ref{eq:theoretical})].
The numerical and theoretical values agreed relatively well.
%This graph resembled the graph of the order parameter of heterogeneous oscillators plotted against the scale parameter of the natural frequencies.
%Indeed, as shown in the following, the graphs were identical in shape.
The continuous transition from the synchronized state to the desynchronized state is observed.

In the literature on phase oscillators, oscillator heterogeneity is often represented by heterogeneous natural frequencies.
The dynamics of oscillator $k$ are
\begin{equation}
  \dot{\phi}_k = \omega_k+\Im[H(t)\exp(-\im\phi_k)]+\sigma(t)\xi_k(t),
\end{equation}
where the natural frequency $\omega_k$ is drawn from the probability density function $g(\omega)$.
In this system, the order parameter of the whole system is defined by
\begin{equation}
  Y(t) = \int_{-\infty}^\infty g(\omega)Z(\omega,t)\di\omega,
\end{equation}
%これはすべての振動子の重心である
which is the center of mass of all oscillators in the system.
In other words, the order parameter of the whole system, $Y(t)$, is the average of the order parameters, $Z(\omega, t)$, of the oscillators with the natural frequency $\omega$.
It has been shown that the Ott--Antonsen ansatz can reduce the dynamics of the order parameter of phase oscillators whose natural frequencies obey a family of rational distribution functions into low-dimensional dynamical equations.
In the most commonly studied version of this system, $g(\omega)$ is the Cauchy distribution,
\begin{equation}
  g(\omega) = \frac{1}{\pi\gamma}\frac{\gamma^2}{(\omega-\omega_0)^2+\gamma^2},
\end{equation}
where $\gamma$ is the scale parameter and $\omega_0$ is the location parameter.
In this case, the Ott--Antonsen low-dimensional dynamics are shown to be given by inserting
\begin{equation}
  Y(t) = Z(\omega_0+\im\gamma,t)
\end{equation}
into Eq.~(\ref{eq:continuousWS}) \cite{Ott2008,Marvel2009,Pikovsky2015}.
Again, it is assumed that the density of the phases of oscillators with frequency $\omega$ follows the wrapped Cauchy distributions and that Eq.~(\ref{eq:Z}) holds for the oscillators with the natural frequency $\omega$.
This assumption approximately holds if the initial phases of a finite number of oscillators are uniformly distributed on $[0,\;2\pi]$.
This results in
\begin{equation}
  \pdiff{Y(t)}{t} = [\im\omega_0-\gamma-\sigma(t)] Y(t)+\frac{H(t)}{2}-\frac{\bar{H}(t)}{2}Y(t)^2. \label{eq:Y}
\end{equation}
Specifically, in globally-coupled phase oscillator systems
\begin{equation}
\dot{\phi}_k = \omega_k+\frac{K}{N}\sum_{j=1}^N\sin(\phi_j-\phi_k)+\sigma\xi_k(t) \label{eq:Kuramoto2}
\end{equation}
with Cauchy-distributed natural frequencies, the mutual interaction is $H(t)=KY(t)$.
Therefore, the dynamics of the order parameter are given by
\begin{equation}
  \pdiff{Y(t)}{t} =  \left(\im\omega_0-\gamma-\sigma+\frac{K}{2}\right)Y(t)-\frac{K}{2}\bar{Y}(t)Y(t)^2.
\end{equation}
Its stable solution is
\begin{equation}
  Y(t) = \begin{cases}
    \sqrt{1-2\frac{\sigma+\gamma}{K}}\exp[\im\omega_0 (t-t_0)] & (\sigma+\gamma\le\frac{K}{2})\\
    0 & \left(\frac{K}{2}<\sigma+\gamma\right)
  \end{cases}. \label{eq:SL2}
\end{equation}
Replacing $Y(t)$, $\omega_0$, and $\gamma+\sigma$ with $Z(\omega,t)$, $\omega$, and $\sigma$ in Eq.~(\ref{eq:SL2}) yields Eq.~(\ref{eq:SL1}).
The macroscopic behavior of the system can be perfectly represented as a function of $\sigma+\gamma$; that is to say, the noise amplitude and the scale parameter of the natural frequency are equivalent in the dynamics of the order parameter.
This means that weak noise and narrowly distributed natural frequencies allow for the synchronization whereas strong noise and widely distributed natural frequencies abolish the synchronization.
%This is the reason why the graphs of the synchronization transition of heterogeneous oscillators and noise-driven oscillator exhibited the same form.

\begin{figure}
  \centering\includegraphics[width=8cm]{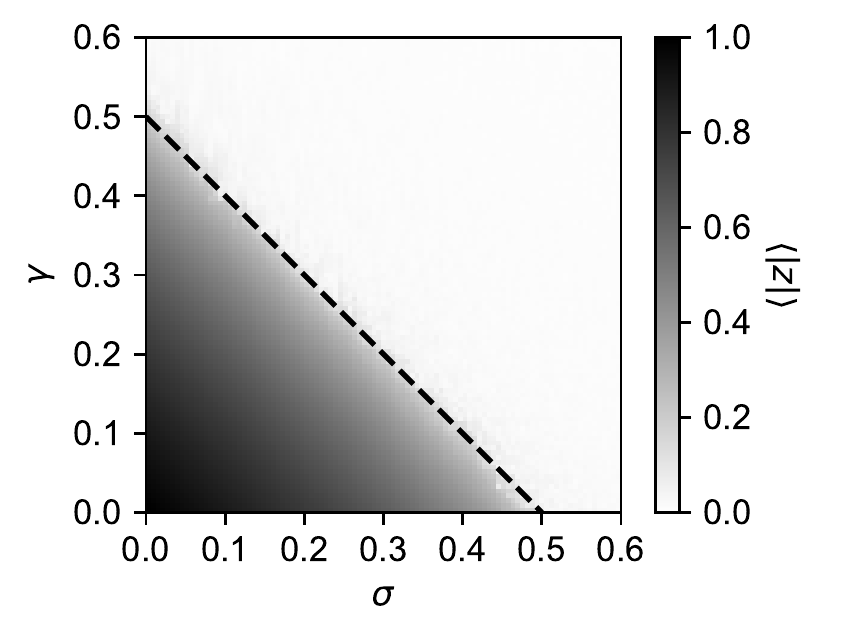}
  \caption{\label{fig:equivalence}
    Numerical results of $\langle|z|\rangle$ for the system of Eq.~(\ref{eq:Kuramoto2}) with $K=1$.
    The theoretically derived boundary between the synchronized state and the desynchronized state is indicated by the dashed line.
}
\end{figure}

To test this analytical result, the simulation of Eq.~(\ref{eq:Kuramoto2}) was performed by the Euler method
\begin{equation}
\phi_k(t+\Delta t) = \phi_k(t)+\Delta t\left(\omega_k + \frac{K}{N}\sum_{j=1}^N \sin[\phi_j(t)-\phi_k(t)]+\sigma\xi_k(t)\right)
\end{equation}
with the same parameter values as in Fig.~\ref{fig:synchronization}.
%オイラー法で
%である．
In Fig.~\ref{fig:equivalence}, the black and white colors correspond to $\langle|z|\rangle=1$ and $0$, respectively.
The dashed line represents the boundary between the synchronized state and the desynchronized state (i.e., $\sigma+\gamma=K/2$).
The figure clearly indicates that the steady-state value of the order parameter is a function of $\sigma+\gamma$.
This supports the equivalence of the noise amplitude and the scale parameter of the natural frequency in the present model.

\section{Discussion}
This paper examined phase oscillator systems driven by Cauchy noise and obtained the low-dimensional description of the dynamics of the order parameter using the Watanabe--Strogatz theory and Ott--Antonsen ansatz.
The low-dimensional dynamics agreed relatively well with the numerical results of a system of a finite number of oscillators.
In the derived low-dimensional dynamics, the scale parameter of the natural frequency, $\gamma$, and the noise amplitude, $\sigma$, were equivalent.
The macroscopic dynamics of the system with heterogeneous natural frequencies were indistinguishable from those of the system driven by Cauchy noise.

The time evolution of the phases of sine-coupled oscillators is described by linear fractional transformations \cite{Marvel2009}.
Linear fractional transformations map the Cauchy distributions to the Cauchy distributions and the wrapped Cauchy distributions on the unit circle to the wrapped Cauchy distributions on the unit circle \cite{McCullagh1992,McCullagh1996}.
As the Cauchy distribution is a stable distribution, it is continued to be obeyed by the oscillators driven by Cauchy noise.
Although oscillators are not microscopically contained in a low-dimensional manifold (because they are driven by independent noise), the trajectories of oscillators driven by sine coupling and Cauchy noise can macroscopically be considered as being confined to a low-dimensional manifold.
Within the framework of the circular cumulant approach \cite{Goldobin2019}, only the first circular cumulant is nonzero in the present model.

The present results shed light on the dynamics of phase oscillators driven by Cauchy noise.
For example, Martens \textit{et al}. investigated the dynamics of phase oscillators whose natural frequencies followed a mixture of two Cauchy distributions \cite{Martens2009}.
The results of the present study combined with those of Martens \textit{et al}. predict the low-dimensional dynamics of Cauchy-noise-driven phase oscillators whose natural frequencies take on one of two values.
The analysis of the conformist and contrarian oscillators \cite{Hong2011} can also be applied to the analysis of noise-driven oscillators.
The present results allow for the reinterpretation of previous analyses on oscillator systems with Cauchy natural frequencies as the analyses on oscillators driven by Cauchy noise.
Whether or not the Gaussian noise facilitates the same type of reinterpretation in certain problem settings is not within the scope of the present research.

The present model assumes that the noise is temporally uncorrelated.
The dynamics of phase oscillators driven by correlated Gaussian noise were previously investigated \cite{Tonjes2010}.
The effect of the correlated Cauchy noise could be investigated by extending the present results.
Because the $1/f$ fluctuation is found in heartbeats \cite{Kobayashi1982} and in the activity of the central nervous system \cite{Novikov1997,Linkenkaer2001}, the analyses of systems driven by temporally correlated noise are likely to find applications in physiology research.

The present study showed that white Cauchy noise and Cauchy-distributed natural frequencies have the same effect on the macroscopic behavior of a specific model.
%In other words, annealed disorder and quenched disorder
In the context of statistical physics, the critical behavior of a $d$-dimensional random field model is related to the critical behavior of a $d-2$-dimensional model without disorder \cite{Parisi1979,Fytas2018,Kaviraj2019}.
The model presented in this study offers a further example of the equivalence of annealed and quenched disorders.
The interplay between annealed and quenched disorders or noise and heterogeneity could be explored further by extending the present model.

\section*{Acknowledgments}
This work was supported by JSPS KAKENHI Grant Number JP19K12184.

%\bibliographystyle{apsrev4-1}
%\bibliography{model}
%merlin.mbs apsrev4-1.bst 2010-07-25 4.21a (PWD, AO, DPC) hacked
%Control: key (0)
%Control: author (72) initials jnrlst
%Control: editor formatted (1) identically to author
%Control: production of article title (-1) disabled
%Control: page (0) single
%Control: year (1) truncated
%Control: production of eprint (0) enabled
%

\end{document}